# LLM-Driven Accessible Interface: A Model-Based Approach

Blessing Jerry [1][0009-0003-3628-3911], Lourdes Moreno [1][0000-0002-9021-2546], Virginia Francisco [2][0000-0002-4492-5633], and Raquel Hervas [2][0000-0003-2900-9992]

[1] Computer Science and Engineering Department, Universidad Carlos III de Madrid, Spain
[2] Facultad de Informática, Universidad Complutense de Madrid, Spain

**Abstract.** The integration of Large Language Models (LLMs) into interactive systems opens new opportunities for adaptive user experiences, yet it also raises challenges regarding accessibility, explainability, and normative compliance. This paper presents an implemented model-driven architecture for generating personalised, multimodal, and accessibility-aligned user interfaces. The approach combines structured user profiles, declarative adaptation rules, and validated prompt templates to refine baseline accessible UI templates that conform to WCAG 2.2 and EN 301 549, tailored to cognitive and sensory support needs. LLMs dynamically transform language complexity, modality, and visual structure, producing outputs such as Plain-Language text, pictograms, and high-contrast layouts aligned with ISO 24495-1 and W3C COGA guidance. A healthcare use case demonstrates how the system generates accessible post-consultation medication instructions tailored to a user profile comprising cognitive disability and hearing impairment. SysML v2 models provide explicit traceability between user needs, adaptation rules, and normative requirements, ensuring explainable and auditable transformations. Grounded in Human-Centered AI (HCAI), the framework incorporates co-design processes and structured feedback mechanisms to guide iterative refinement and support trustworthy generative behaviour.

## 1 Introduction

The growing integration of generative AI, particularly Large Language Models (LLMs), is reshaping how user interfaces (UIs) can adapt to diverse users by generating content and structures dynamically in real time. While this shift offers new opportunities for personalization, it also raises challenges regarding inclusivity, explainability, and normative alignment, especially in sensitive domains such as health, education, and public services.

Current generative systems lack systematic mechanisms to ensure that outputs are accessible for people with disabilities or to trace how user characteristics influence generated content and interface structures. These limitations are particularly problematic in light of regulatory frameworks such as the European Accessibility Act and standards including WCAG 2.2 [1] and EN 301 549 [2].

This paper presents an implemented modular architecture and model-driven approach that supports the generation of personalised, multimodal, and accessibility-aligned user interfaces using LLMs. The system integrates structured user profiles, declarative adaptation rules, and generative components to produce interfaces tailored to cognitive and sensory needs. All UI templates conform to WCAG 2.2 and EN 301 549 by default, while the UserProfile specifies additional refinements (e.g., Plain-Language, stepwise structuring, pictograms) derived from recognised accessibility guidelines. The architecture is instantiated through SysML v2 models and adaptive UI templates that ensure traceability and explainability.

To demonstrate feasibility, the approach is applied to a healthcare scenario in which a teleconsultation platform is extended with a generative module that produces accessible post-consultation medication instructions. The system adapts outputs to a user profile combining cognitive disability and hearing impairment, and the associated generative workflow is modelled end-to-end in SysML v2. While a full user evaluation is planned as future work, the current prototype incorporates normative constraints, structured feedback mechanisms, and reusable templates that support auditable, standards-aligned adaptation.

The remainder of this paper is structured as follows. Section 2 reviews work on generative personalisation, accessibility, human-centred design, and model-driven engineering. Section 3 introduces the proposed architecture, design principles, and conceptual metamodel. Section 4 presents a healthcare case study with SysML v2 models, adaptation rules, and UI templates. Section 5 discusses findings, limitations, and future directions.

## 2      Related Work

We organize the related work into five areas that contextualize our contribution: accessibility, generative UI personalization, human-centred design, traceability, and model-driven engineering.

### 2.1    Accessible User Interface Design

Accessible interface design must incorporate accessibility from the outset; delayed integration often results in inadequate solutions. Standards such as WCAG 2.2, EN 301 549, ISO 24495-1 (Plain language Governing principles and guidelines), and W3C COGA guidelines [3] define criteria ensuring interfaces are perceivable, operable, understandable, and robust. Effective accessible systems utilize multimodal adaptations (text, images, video, audio descriptions, and pictograms) and support assistive technologies such as screen readers, keyboard navigation, and high-contrast visuals.

Recent research extends these principles through adaptive and generative approaches that personalize user experiences. [4] assessed generative tools against WCAG 2.2 benchmarks and revealed persistent gaps in automated accessibility verification. Building on this, [5] reported that many systems still fail to meet the needs of users with visual, motor, and cognitive disabilities. [6] explored LLM-based techniques for simplifying complex text into Easy-to-Read formats, advancing cognitive accessibility, and [7] proposed adaptive frameworks embedding accessibility features directly interface generation. In addition, [8] demonstrated how generative systems can personalize 3D environments for users who are blind and deaf.

However, despite these advances, accessibility standards and multimodal adaptations remain weakly integrated into generative pipelines, and most systems lack mechanisms for verifiable compliance.

### 2.2 Human-Centred and Participatory Design

Human-Centred AI (HCAI) emphasizes iterative co-design, user participation, and transparency, which are essential when designing for diverse populations (ISO 9241-210). [9] co-created an LLM-based personal health assistant with older adults, explicitly addressing vocabulary preferences and explainability. [10] employed a dual-mode chatbot for retirement communities, improving usability through voice and simplified text interactions tailored to accessibility needs. Similar participatory approaches are evident in work by [11–15], who collaborated with users across sensory, cognitive, and neurodivergent contexts to inform adaptive interface behavior. Other studies, including those by [16] and [17], highlight how behavioral adaptation and explainability reinforce user agency and transparency.

These studies consistently emphasize the importance of traceability between user-stated requirements and generative system behaviors, a foundational principle for developing accountable and adaptive AI. Participatory methods not only improve technical outcomes but also foster user trust and alignment with real-world needs, particularly in sensitive domains such as health, education, and assistive technologies. Collectively, this body of work shows that involving users throughout the adaptation process is crucial for developing inclusive and trustworthy human-centered AI systems. However, these approaches seldom connect participatory insights with formal modelling or structured visual adaptation, leaving normative compliance and traceable adaptation logic insufficiently addressed.

### 2.3 Generative AI for User Interface Personalization

Generative User Interfaces (GenUIs) powered by Large Language Models (LLMs) dynamically produce adaptive content based on user profiles, leveraging technologies such as GPT-driven renderers. [14] developed an assistive agent that adapts interactions between images and dialogues according to user strengths. Nevertheless, critical analyses indicate that most GenUIs neglect established accessibility standards [4, 5]. Recent advancements aim to address these shortcomings by incorporating user feedback through reinforcement learning frameworks [18] and prompt engineering for adaptive learning paths [19]. [20, 21] introduced PromptInfuser, a plugin that tightly couples UI mock-ups with LLM prompts to support intent-driven, semi-functional prototyping. While not directly focused on accessibility, it demonstrates how prompt-driven generative tools can improve interface fidelity and better align system behaviour with designer intent. Complementing these design-stage developments from a runtime perspective, [8] proposed SceneGenA11y, a system that personalizes 3D environments for accessibility, demonstrating real-time adaptation to diverse sensory needs. Furthermore, [7] advanced adaptive frameworks that integrate accessibility into self-adaptive systems, emphasizing continuous usability enhancement. However, despite these advances, current GenUI systems still lack mechanisms for enforcing accessibility constraints or providing verifiable, traceable adaptation decisions.

### 2.4    Traceability and Explainability in Adaptive Systems

Traceability and explainability are essential for maintaining trust and accountability in adaptive systems. As model-driven and generative methods increasingly inform accessibility design, documenting how adaptation decisions are made has become a central concern. Transparent design practices enable users to understand how their data and interactions influence system behavior, and research on explainable interfaces reinforces this need for clarity. [17] identified transparency and traceability as crucial for explainable interfaces, noting that many adaptive systems lack clear reasoning trails. Interpretable feedback mechanisms can enhance trust in AI systems; however, as discussed by [22], explainability often diminishes as personalization increases. [23] highlighted that generative accessibility systems provide limited insight into how adaptations follow accessibility principles. Similarly, [4] emphasized that current generative pipelines rarely capture verifiable evidence of compliance, restricting accountability. Building on this, [17] advanced model-driven accessibility automation but offered minimal human oversight or iterative feedback integration. These studies reveal persistent gaps in transparent adaptation logic and human-in-the-loop participation. Addressing these gaps is key to ensuring that adaptive accessibility systems remain auditable, interpretable, and responsive to diverse user needs.

### 2.5    Model-Driven Engineering for Accessibility

Model-Driven Development (MDD/MDE) employs abstract models for systematic interface generation, supporting adaptive and personalized user experiences. Early initiatives, including CAMELEON, SUPPLE, MyUI, and AdaptForge, established structured multimodal adaptations, particularly for users with disabilities or age-related conditions. Recent integrations of LLMs into MDE pipelines highlight enhanced personalization potential, exemplified by UICoder [24], which generates valid HTML/CSS through iterative user feedback, and GPT-driven UI collaboration frameworks [25]. Within this space, [7] introduced an adaptive framework that incorporates accessibility features within model-driven systems, allowing interfaces to adjust dynamically to user needs. [8] extended this approach by combining model-based methods with generative reasoning from large language models to improve interface synthesis. Complementary evaluations by [4] assessed generative systems under WCAG 2.2 benchmarks, emphasizing the importance of systematic validation in accessible model-driven development. [6] explored cognitive accessibility by proposing model-informed simplification strategies supported by language models. [15] further demonstrated how generative modelling can personalize complex 3D environments, illustrating the potential for integrating runtime adaptation with model-driven logic. Despite these advances, few approaches combine LLM capabilities with comprehensive accessibility adherence and rigorous, verifiable traceability.

Thus, significant gaps persist across these domains: user interfaces often neglect multimodal adaptations aligned explicitly with regulatory standards; GenUIs insufficiently address accessibility; HCAI studies rarely integrate structured modelling; and MDD/MDE approaches seldom utilize generative AI fully aligned with normative compliance. This paper presents these gaps by proposing a standards-aligned, model-based framework leveraging LLMs to generate adaptive, multimodal, and accessible user interfaces with robust traceability, normative alignment, and meaningful user involvement.

## 3   Conceptual Approach for Adaptive Accessible User Interface with Generative AI

This section presents the conceptual foundation of the proposed approach to generating personalized and accessible user interfaces using LLMs. The proposed system combines MDD, user-centric adaptation, and regulatory alignment to deliver traceable and explainable generative outputs. The section introduces the core design principles, a layered architecture, key system components, and a formal conceptual metamodel.

### 3.1   Design Principles

The framework addresses limitations identified in prior work on generative UIs, accessibility modelling, and human-centred AI by integrating normative constraints, modular design, semantic personalization, and participatory processes within a unified architecture (Figure 1). It adheres to accessibility standards such as WCAG 2.2 and EN 301 549 and incorporates additional requirements from ISO 24495-1 (Plain Language) and the W3C COGA guidelines for cognitive accessibility.

- **Normative Alignment and Traceability**: Beyond compliance, the system emphasizes traceability between user profiles, adaptation rules, and generated outputs. Each transformation is informed by a structured *UserProfile* and linked to corresponding adaptation logic, enabling behaviour to be traced back to both normative justifications and user needs. SysML v2 modelling provides the formal constructs required to represent these relationships explicitly and consistently.
- **Modularity and Semantic Personalization**: the modular architecture allows the reuse of components and the addition of new adaptation functions without altering the core engine. Semantic personalization leverages LLMs and prompt engineering, adapting content in real-time to match user preferences.
- **Human-Centred Participation**: the framework incorporates participatory processes by involving people with disabilities in design, validation, and

- feedback. Their input refines adaptation logic and ensures that the generative behaviour reflects real-world needs.
- **Human Oversight and Governance**: Human supervision is incorporated following a Human-on-the-Loop (HoTL) paradigm. Users and domain experts can review, adjust, or reject adaptations at key points, and structured feedback is linked to requirements and component identifiers to support accountable updates. HoTL oversight is reinforced through versioned models and auditable logs that allow supervisors to inspect and correct adaptation decisions ex post, while runtime generation operates autonomously under normative constraints.
- **Safety, Quality, and Privacy Controls**: Technical robustness is supported by automatic checkpoints for readability, semantic fidelity, factual consistency, and perceived clarity. Readability follows Plain-Language guidance, while semantic and factual checks mitigate meaning drift and unsupported claims. These checkpoints function as validation gates: outputs that fail a check trigger a controlled regeneration cycle or are escalated for Human-on-the-Loop (HoTL) review, ensuring that unsuitable adaptations are corrected before presentation to the user. Gates run before content is shown and are re-evaluated whenever text changes, reducing hallucinations and maintaining alignment with normative constraints.

  Privacy and data governance follow the principle of data minimization. User profiles store only the capabilities required for adaptation, while feedback and compliance records use anonymised or simulated data when appropriate. Only non-identifying references (e.g., requirement or component identifiers) are persisted.
- **Transparency and Compliance**: Transparency is achieved through SysML v2 trace and verify links, serialized evidence, and machine-readable reports documenting how outputs were generated. Accessibility is treated as a dimension of fairness, with explicit support for cognitive, visual, and motor needs. By binding adaptations to explicit models, enforcing standards at generation time, and preserving human oversight with traceable evidence, the approach ensures compliance with WCAG 2.2, EN 301 549, ISO 24495-1, and the W3C COGA guidelines, and aligns with the requirements of the European Trustworthy AI framework (EC HLEG on AI, 2019).

## 3.2   Layered Architecture Overview

The proposed architecture comprises five layers, each supporting a distinct function in the adaptive generation workflow (see Figure 1). The Conceptual Layer defines the underlying interaction metamodel, capturing users, goals, tasks, and accessibility needs. The Adaptation Layer formalises the mapping between user-profile conditions and content or interface transformations through declarative semantic and logical rules aligned with accessibility standards. The Generative Layer operationalises these rules through LLM-based transformations and prompt engineering, enabling content simplification, modality adaptation, and structural reorganisation. The UI Rendering Layer produces the adapted interface on the target platform and enforces technical accessibility requirements. Finally, the Evaluation and Co-Design Layer collects user feedback and supports participatory validation, enabling iterative refinement of rules, templates, and preferences to ensure the system remains responsive to user needs.

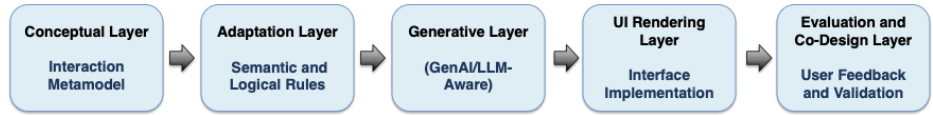

**Figure 1.** Proposed Architecture (Layer Model)

### 3.3 System Modelling and Adaptation Logic

To ensure structured, auditable, and explainable adaptation, a conceptual metamodel is defined to capture the core elements of the system and their interrelations (see Figure 2). The metamodel comprises six main entities: `UserProfile`, `DomainInput`, `AdaptationRule`, `UIComponent`, `OutputModality`, and `Feedback`, together with a Normative Accessibility Requirements element that enables traceable alignment with accessibility standards.

- The `UserProfile` encapsulates user attributes, accessibility needs, and preferred interaction modalities, and refines the Normative Accessibility Requirements, selecting the clauses of WCAG 2.2, EN 301 549, ISO 24495-1, W3C COGA, or the Trustworthy AI Guidelines that apply to the user.
- `DomainInput` represents the raw content to be transformed (e.g., a post-consultation summary or task instruction) and triggers the adaptation process.
- The adaptation logic is captured in the `AdaptationRule` entity. Each rule links profile-derived conditions (e.g., "requires cognitive support") with specific transformations (e.g., Plain-Language simplification or the addition of pictograms). Rules include identifiers, conditions, targets, priorities, and associated prompts, and they inform both the generative transformation and subsequent layout decisions.
- `UIComponent` constitute the building blocks of the interface. They may be atomic (e.g., buttons or labels) or composite (e.g., containers), and include component identifiers, types, content, and requirement references. Components are adapted according to the transformation logic defined in the active rules.
- `OutputModality` specifies the delivery format of adapted content, text, audio, pictogram, or video, and is selected based on user needs and available transformations.
- The `Feedback` entity captures comprehension ratings, navigation behaviour, and user comments. These inputs inform prompt tuning, rule refinement, and profile updates, enabling continuous improvement in adaptive behaviour.

The generative adaptation process combines user profiling, rule-based logic, prompt-driven transformation, and multimodal rendering. Traceability is achieved by linking `AdaptationRule` and `OutputModality` elements to normative requirements, enabling verification of compliance with standards such as WCAG, EN 301 549, and ISO 24495-1. A curated repository of prompt templates, co-designed with users with disabilities and domain experts, reduces bias and enhances accessibility. Evaluation integrates automated NLP metrics, user validation, UX assessment, and modality effectiveness to measure trust, clarity, and explainability, while human-in-the-loop feedback ensures robust personalization.

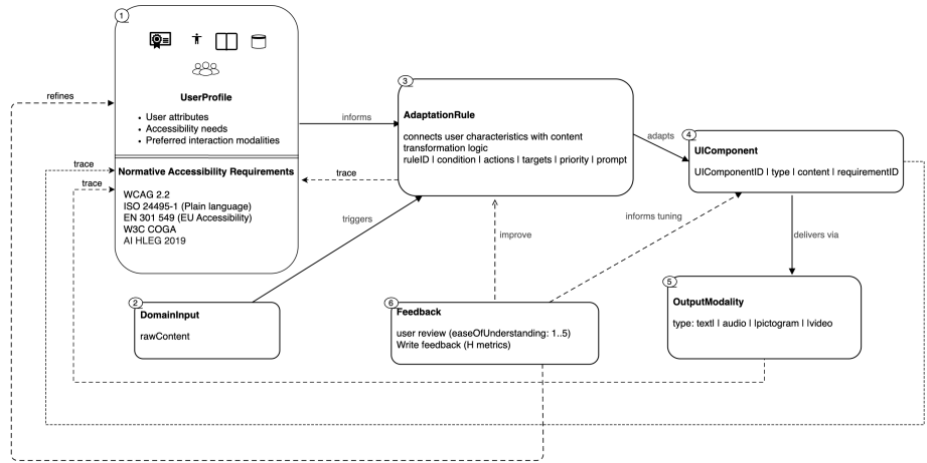

**Figure 2.** Conceptual metamodel showing the six core entities and their relationships

## 4   Case Study: Post-Consultation Medical Interface

Building on the conceptual metamodel and layered architecture introduced in Section 3, this section demonstrates how the adaptive generative framework is instantiated in a real medical communication scenario. The case study operationalises the user-profiling constructs, adaptation rules, and multimodal generation mechanisms to support users with cognitive and sensory disabilities in understanding post-consultation information.

### 4.1   Context and Use case Scenario

This case study examines the application of the proposed adaptive generative architecture to accessible medical communication. It builds upon Access2meet [26, 27], a videoconferencing platform designed to support accessibility in remote healthcare consultations. The platform was developed in alignment with WCAG 2.1 and EN 301 549 requirements and accommodates users with cognitive, visual, auditory, and motor impairments. It provides real-time accessibility services, including automatic subtitling, plain-language reformulation, and alternative input and output modalities, and integrates cognitive-accessibility UI patterns co-designed with users, caregivers, and healthcare professionals.

The scenario focuses on a patient receiving post-consultation medication instructions after a remote appointment. In this use case, the user presents both mild cognitive disability and hearing impairment and therefore requires information that is simplified, visually oriented, and structured into clear steps. The original medical note contains domain-specific terminology and syntactic complexity that hinder comprehension. In this scenario, content adaptation is guided by the accessibility requirements encoded in the user's `UserProfile`, which references normative frameworks such as WCAG 2.2, EN 301 549, ISO 24495-1 (Plain Language), and W3C COGA. These requirements inform the derivation of concrete adaptation rules, detailed in Section 4.3.1, and serve as the basis for generating an accessible interface rendered using a React-based schema.

This scenario provides a concrete and illustrative setting to demonstrate how the model-driven architecture supports clarity, autonomy, and accessibility in sensitive healthcare communication.

### 4.2   System Modelling

To ensure formal specification, explainability, and traceability, the architecture is modelled in SysML v2 by instantiating the conceptual metamodel introduced in Section 3. The modelling artefacts operationalise the abstract entities of the metamodel and apply them to the post-consultation scenario described in Section 4.1. The resulting models cover structural, behavioural, and requirements views aligned with the adaptive workflow.

**Structural View.** Figure 3 presents a simplified Block Definition Diagram (BDD) capturing the main architectural components:

- `AdaptiveInterfaceSystem`: top-level block encapsulating the adaptive architecture.
- `UserProfile`: stores accessibility parameters and interaction preferences.
- `MedicalPrescription`: domain input containing the consultation text.
- `GenAIEngine`: performs prompt-based simplification and activates adaptation rules.
- `UIAdapter`: matches transformed content to appropriate layouts and output modalities.
- `UIRenderer`: assembles and displays the adapted UI.
- `FeedbackCollector`: captures indicators of comprehension, interaction behaviour, and accessibility feedback.

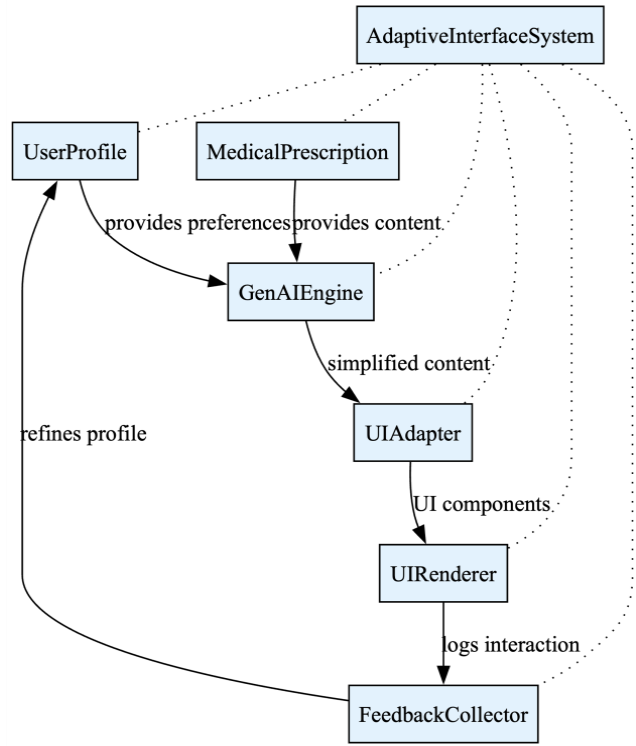

**Figure 3.** Block Definition Diagram

These blocks are linked through typed associations and data-flow connections. Each element corresponds directly to its counterpart in the conceptual metamodel (Section 3), ensuring traceability from abstract models to implemented modules.

**Behavioural View.** The adaptation workflow is captured through:

- an **Activity Diagram** detailing the sequence from retrieving the user profile, applying transformation rules to domain input, generating adapted content via the `GenAIEngine`, selecting appropriate modalities with the `UIAdapter`, and rendering the final interface.
- a **State Machine Diagram** defining user interaction states (Reading, Navigating Steps, Requesting Help, Completing Task), enabling dynamic UI behaviour based on real-time interaction.

**Requirements View.** A **Requirement Diagram** defines alignment with accessibility standards by mapping system functions to normative criteria:
- Plain-Language compliance (`REQ-PL-01`, ISO 24495-1; WCAG 3.1.5; COGA)

- Visual contrast and avoidance of colour-only cues (`REQ-WCAG-01`, WCAG 1.4.1-1.4.3)
- Multimodal presentation combining text and pictograms (`REQ-MOD-02`, WCAG 1.1.1; COGA)
- User-feedback logging for iterative refinement (`REQ-FB-01`)

These requirements are linked to system components through satisfy and trace relationships, providing formal auditability and verifiable compliance.

Grounded in formal modelling, the system supports the generation of personalised, multimodal interfaces adapted to cognitive and sensory accessibility needs. Unlike prior LLM-based systems focused on purely textual or conversational outputs, this framework combines structured UI generation with explicit normative alignment.

### 4.3    Overview of the Generative Adaptation Pipeline

The generative adaptation workflow operationalises the model-driven architecture by combining rule-based logic with LLM-driven transformations. The process takes as input the accessibility requirements encoded in the user's `UserProfile`, the medical content to be adapted, and the normative references defined in the conceptual metamodel. Adaptation is performed through a sequence of interpretable steps that ensure traceability, explainability, and alignment with accessibility standards. The workflow proceeds as follows:

- **Profile interpretation**: The system interprets accessibility needs specified in the `UserProfile`. These needs are associated with normative references from WCAG 2.2, EN 301 549, ISO 24495-1, and W3C COGA.
- **Rule activation**: The interpreted needs activate a subset of declarative `AdaptationRules` (see Section 4.3.1), each of which links a condition to a transformation.
- **Prompt construction**: A structured prompt template is instantiated with the domain input, the active rules, and profile-specific parameters.
- **Generative transformation**: The `GenAIEngine` applies the instantiated prompt using an LLM fine-tuned on Plain-Language and Easy-to-Read corpora, producing simplified text, pictogram-aligned descriptions, and modality-specific representations.
- **UI adaptation**: The `UIAdapter` selects layouts, contrast modes, and modalities based on the transformed content and user requirements, ensuring technical accessibility constraints are satisfied.
- **Rendering and feedback**: The `UIRenderer` produces the adapted interface, and the `FeedbackCollector` records user interaction and comprehension data for subsequent refinement.

This workflow ensures that each generated output can be traced back to: (i) a user need, (ii) a declared rule, (iii) a normative requirement, and (iv) a specific step in the generative process.

This operationalises the formal modelling introduced in Section 4.2 and provides an explainable, auditable adaptation pipeline suitable for sensitive medical communication.

**Derived Accessibility Requirements.** All UI templates produced by the framework conform by default to WCAG 2.2 and EN 301 549. The UserProfile does not "enable" accessibility; instead, it refines and personalises additional cognitive- and sensory-support patterns (e.g., Plain Language, stepwise structuring, pictograms). These complements operate on top of the baseline accessibility guarantees.

To provide traceable, standards-aligned adaptation, the accessibility needs of users with cognitive disabilities and hearing impairments are formalised into a set of Derived Accessibility Requirements (DAR), summarised in Table 1. Each DAR establishes a link between a user need, an adaptation requirement, a transformation rule, and the corresponding normative reference.

Table 1. Derived Accessibility Requirements (DAR) linked to adaptation rules and normative standards

| User Need | Derived Requirement (DAR) | Adaptation Rule | Normative Reference |
|---|---|---|---|
| Cognitive disability | DAR-01: Content shall be reformulated in Plain Language | simplifyText() | W3C COGA (language & structure); Guideline 3.1 Readable |
| Cognitive disability | DAR-02: Instructions shall be structured into step-wise segments. | structureAsSteps() | WCAG 2.4.6; COGA "Sequential Steps" |
| Cognitive disability | DAR-03: Key medical actions shall include pictograms. | attachPictograms() | COGA "Reinforce Meaning" |
| Hearing impairment | DAR-04: Auditory information shall be replaced with visual alternatives. | disableAudio(), enableVisualAlerts() | WCAG 1.2.1; WCAG 1.4.1 |
| Hearing impairment | DAR-05: Critical information shall use a high-contrast visual presentation. | applyHighContrast() | WCAG 1.4.3 |
| Motor/cognitive load | DAR-06: Interaction elements shall use large touch targets. | renderLargeTargets() | WCAG 2.5.5 |
| General clarity | DAR-07: Content shall use short phrases and bullet-point structure. | simplifyStructure() | COGA "Reduce Cognitive Load" |

These requirements are encoded in a SysML Requirement Diagram and linked to system components through trace and satisfy relationships, ensuring a verifiable alignment between high-level accessibility constraints and concrete implementation decisions.

**Quality Gates and Controlled Regeneration.** The generative adaptation workflow operationalises the metamodel by applying user-derived accessibility requirements (DAR) to the domain input in a traceable and standards-aligned manner. The workflow begins with the interpretation of the UserProfile, which activates the relevant Derived Accessibility Requirements (Section 4.3.1). Each DAR corresponds to one or more declarative AdaptationRules, specifying both the condition and the transformation to be applied.

These active rules inform a structured prompt template instantiated with the consultation text, rule parameters, and user-specific preferences. The GenAIEngine executes the prompt using an LLM specialised in Plain-Language and Easy-to-Read content, producing simplified text and pictogram-enhanced descriptions consistent with the user's cognitive and sensory accessibility needs.

After the generative transformation, the system applies automatic quality checkpoints that assess readability, semantic fidelity, and factual consistency. These checkpoints act as validation gates: if the generated output deviates from Plain-Language guidance, alters the intended meaning, or introduces unsupported claims, the system triggers a regeneration cycle or escalates the output for Human-on-the-Loop (HoTL) review. This ensures that only accessible, reliable, and normatively aligned content proceeds to UI adaptation and rendering.

The UIAdapter selects the appropriate output modalities, such as Plain-Language text, pictograms, high-contrast layouts, or visual-only rendering, based on the transformations generated by the engine and the user's profile. The UIRenderer composes the interface using validated layout templates, ensuring conformance with WCAG 2.2, EN 301 549, and ISO 24495-1.

Throughout the workflow, transformation steps, rule activations, and modality selections are logged via the trace links defined in the SysML v2 models. This ensures that every output can be traced back to (i) a user need, (ii) an activated rule, and (iii) its associated normative requirement. The process therefore provides an explainable, auditable, and model-driven adaptation pipeline suitable for sensitive healthcare communication.

### 4.4  Generated Interface Mockups and Adaptation Templates

To ground the adaptation workflow in concrete UI realisations, we developed interface templates and mockups that reflect the multimodal outputs produced by the generative layer. These templates follow the structural and behavioural constraints defined by the SysML v2 models and integrate accessibility guidelines from WCAG 2.2, EN 301 549, ISO 24495-1, and W3C COGA.

The adapted interface uses a step-by-step visual layout, with each instruction presented as a discrete, numbered block. Plain-Language text is reinforced with pictograms aligned with a medical semantic map, ensuring recognisable representation of key actions (e.g., medication intake, dosage intervals). High-contrast buttons facilitate navigation, and large, well-spaced targets support users with motor and cognitive load constraints.

The prompt template used by the GenAIEngine is populated with user-specific attributes and domain input. A simplified example includes placeholders such as:

```
[Instruction]: Simplify this medical note using Plain-Language and add pictograms.
[UserProfile]: {cognitiveSupport: true, auditoryExclusion: true}
[InputText]: "You should take Ibuprofen 400mg every 8 hours unless you experience gastric discomfort."
```

This prompt triggers the production of three outputs: (i) a Plain-Language version of the instruction, (ii) a pictogram-enriched version, and (iii) a step-wise structured flow.

These outputs are mapped to UI templates defined in the Technical UI Layer, which ensures compatibility with the React-based rendering schema used in the implementation. The resulting interface (see Figure 4) provides a personalised, comprehensible, and normatively aligned interaction experience tailored to the needs of users with cognitive and sensory disabilities.

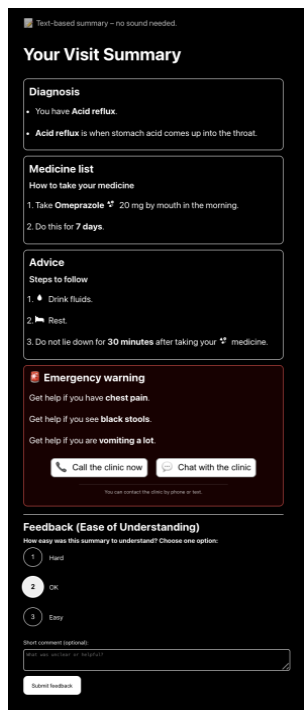

**Figure 4.** Adaptive Consultation Interface. The generated adaptive UI displays pictogram-enhanced medical information and a comprehension feedback scale for accessibility evaluation.

### 4.5  Model Implementation

The case study demonstrates how the conceptual model and layered architecture are implemented through a SysML-driven adaptation pipeline. Figure 5 illustrates a structural subset of the implementation model, connecting user needs, content transformation, and multimodal UI generation through typed associations and traceable rule activation paths. These artefacts validate the architectural consistency of the approach and provide a foundation for repeatable, auditable adaptive behaviour.

All adaptations executed by the generative engine are logged as machine-readable traceability records, linking the applied rules, user requirements, and normative references involved in each transformation. This enables transparent inspection of how content was simplified, how pictograms were selected, and how technical UI constraints were enforced. Such traceability supports the principles of explainability, accountability, and transparency defined by the AI HLEG (2019), ensuring that generative behaviours remain auditable in sensitive healthcare contexts.

Ethical safeguards complement the technical implementation. These include co-designed prompt templates, explicit constraints preventing biased or exclusionary outputs, and multimodal adaptation strategies that respect user autonomy and cognitive load. The pipeline is intentionally extensible: new rules, modalities, and constraints can be added without modifying the core architecture, due to the modular structure formalised in the SysML v2 metamodel.

As part of ongoing work, we plan to conduct a user-in-the-loop evaluation involving individuals with cognitive and sensory disabilities. This evaluation will compare different generative variants, such as alternative simplification strategies or layout structures and assess user trust, clarity, and preference. These insights will be fed back into the rule set, prompting refinements to both the prompt templates and the formal modelling artefacts, reinforcing the system's iterative, human-centred nature.

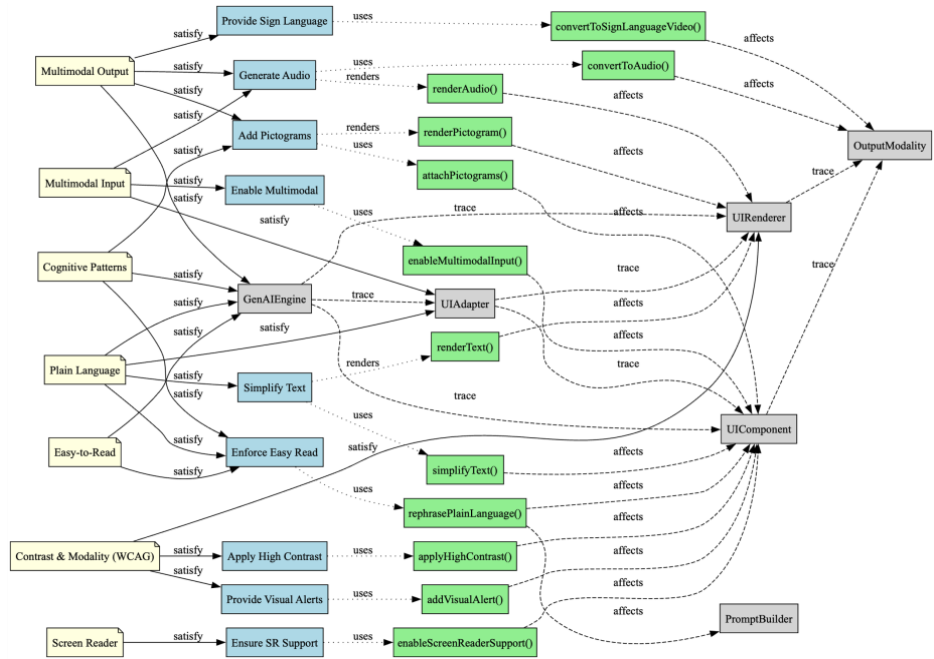

**Figure 5.** Layered Model of User-Centred Adaptive UI Design

## 5    Discussion and Conclusion

This work has presented an adaptive generative framework that integrates model-driven engineering, normative alignment, and large language models to produce accessible, personalised user interfaces for users with cognitive and sensory disabilities. The approach addresses a key gap in existing LLM-based systems, the lack of structured, standards-aligned adaptation, by grounding generative behaviour in a conceptual metamodel and its corresponding SysML v2 instantiations.

A central contribution of the framework is the use of SysML v2 as a mechanism for explainability, traceability, and verifiable alignment with accessibility standards. User needs, transformation rules, adaptation workflows, and normative requirements are represented within a unified model, enabling auditable links between accessibility constraints, the rules they activate, and the generative outputs they produce. This supports transparent reasoning about interface adaptations and allows controlled evolution of the system as standards or user requirements change.

The architecture operationalises adaptation through a generative workflow that combines declarative rule-based logic with LLM-driven transformations. The GenAI Engine performs Plain-Language simplification and multimodal content generation, while the UIAdapter selects layouts, modalities, and contrast settings in accordance with cognitive-accessibility principles. The healthcare case study demonstrates how this combination of structured modelling, normative requirements, and generative templates can produce stepwise instructions, pictogram-enhanced content, and high-contrast interfaces aligned with WCAG 2.2, EN 301 549, ISO 24495-1, and W3C COGA guidance.

The case study also highlights the relevance of Derived Accessibility Requirements (DARs), which translate normative criteria into concrete, activable adaptation rules. These DARs provide a systematic interpretation of user profiles and a predictable, reproducible basis for transformation logic. The governance layer follows a Human-on-the-Loop (HoTL) paradigm, enabling experts to review and refine generative behaviour through structured feedback and versioned model artefacts while allowing runtime adaptation to proceed autonomously under normative constraints.

Despite these contributions, limitations remain. The current prototype relies on prompt-based control of generative models, which may introduce variability despite the presence of constraints. The workflow does not yet incorporate runtime uncertainty estimation or automated verification beyond rule-based checks. Future work will extend evaluation to larger and more diverse user groups, integrate model-based simulation for safety analysis, and explore automated synthesis of adaptation rules from empirical evidence. A user-in-the-loop study is planned to assess effects on trust, comprehension, and real-world accessibility outcomes.

Overall, the results demonstrate how model-driven engineering and generative AI can be combined within a normative, auditable framework to support inclusive interface generation. By unifying formal modelling, declarative adaptation logic, and human-centred principles, the architecture provides a reproducible and standards-aligned foundation for future developments in accessible generative systems.

**Acknowledgments.** This work has also been supported by grants PID2023-148577OB-C21 (Human-Centered AI: User-Driven Adapted Language Models-HUMAN_AI) and PID2023-148577OB-C22 (Human-Centered AI: User-Driven Adaptative Interfaces-HumanAI_UI) funded by MICIU/AEI/10.13039/501100011033 and by FEDER/UE.

## References

1. W3C WAI: Web Content Accessibility Guidelines (WCAG) 2.2. (2023).
2. HF: ETSI EN 301 549 - V3.2.1 - Accessibility requirements for ICT products and services. (2021).
3. W3C WAI: Making Content Usable for People with Cognitive and Learning Disabilities (COGA). (2021).
4. Acosta-Vargas, P., Salvador-Acosta, B., Novillo-Villegas, S., Sarantis, D., Salvador-Ullauri, L.: Generative Artificial Intelligence and Web Accessibility: Towards an Inclusive and Sustainable Future. Emerging Science Journal. 8, 1602–1621 (2024). https://doi.org/10.28991/ESJ-2024-08-04-021.
5. Alshaigy, B., Grande, V.: Forgotten Again: Addressing Accessibility Challenges of Generative AI Tools for People with Disabilities. ACM International Conference Proceeding Series. (2024). https://doi.org/10.1145/3677045.3685493.

6.  Ledoyen, F., Dias, G., Lechervy, A., Pantin, J., Maurel, F., Chahir, Y., Gouzonnat, E., Berthelot, M., Moravac, S., Altinier, A., Khairalla, A.: Inclusive Easy-to-Read Generation for Individuals with Cognitive Impairments. (2025).
7.  Wickramathilaka, S., Grundy, J., Madampe, K., Haggag, O.: Adaptive and accessible user interfaces for seniors through model-driven engineering. Automated Software Engineering. 32, 1–42 (2025). https://doi.org/10.1007/S10515-025-00547-Z/TABLES/3.
8.  Cao, Y., Jiang, P., Xia, H.: Generative and Malleable User Interfaces with Generative and Evolving Task-Driven Data Model. Conference on Human Factors in Computing Systems - Proceedings . (2025). https://doi.org/10.1145/3706598.3713285/SUPPL_FILE/PN7563-TALK-VIDEO.MP4.
9.  Mahmood, A., Cao, S., Stiber, M., Antony, V.N., Huang, C.-M.: Voice Assistants for Health Self-Management: Designing for and with Older Adults. Conference on Human Factors in Computing Systems - Proceedings . 1, (2025). https://doi.org/10.1145/3706598.3713839.
10. Li, L.X., Chung, R., Chen, F., Zeng, W., Jeon, Y., Zaslavsky, O.: Learning from Elders: Making an LLM-powered Chatbot for Retirement Communities more Accessible through User-centered Design. (2025). https://doi.org/10.5281/zenodo.15292697.
11. Ferreira, R. dos S., Castro, T.H.C. de, Ferreira, R. dos S., Castro, T.H.C. de: Participatory and Inclusive Design Models from the Perspective of Universal Design for Children with Autism: A Systematic Review. Education Sciences 2024, Vol. 14,. 14, (2024). https://doi.org/10.3390/EDUCSCI14060613.
12. Zhang, Z., Thompson, J.R., Shah, A., Agrawal, M., Sarikaya, A., Wobbrock, J.O., Cutrell, E., Lee, B.: ChartA11y: Designing Accessible Touch Experiences of Visualizations with Blind Smartphone Users. ASSETS 2024 - Proceedings of the 26th International ACM SIGACCESS Conference on Computers and Accessibility. 1, 2024 (2024). https://doi.org/10.1145/3663548.3675611.
13. Carik, B., Izaac, V., Ding, X., Scarpa, A., Rho, E.: Reimagining Support: Exploring Autistic Individuals' Visions for AI in Coping with Negative Self-Talk. CHI Conference on Human Factors in Computing Systems (CHI '25), April 26-May 1, 2025, Yokohama, Japan. 1, (2025). https://doi.org/10.1145/3706598.3714287.
14. Rajagopal, A., Nirmala, V., Jebadurai, I.J., Vedamanickam, A.M., Kumar, P.U.: Design of Generative Multimodal AI Agents to Enable Persons with Learning Disability. ACM International Conference Proceeding Series. 259–271 (2023). https://doi.org/10.1145/3610661.3617514.
15. Cao, Y., HE, Y., Chen, Y., Chen, M., You, S., Qiu, Y., Liu, M., Luo, C., Zheng, C., Tong, X., Liang, J., Gong, J.: Designing LLM-simulated Immersive Spaces to Enhance Autistic Children's Social Affordances Understanding. Proceedings of the 30th International Conference on Intelligent User Interfaces. 1, 519–537 (2025). https://doi.org/10.1145/3708359.3712142.
16. Swaroop, S., Buçinca, Z., Gajos, K.Z., Doshi-Velez, F.: Personalising AI Assistance Based on Overreliance Rate in AI-Assisted Decision Making.

International Conference on Intelligent User Interfaces, Proceedings IUI. 1107–1122 (2025). https://doi.org/10.1145/3708359.3712128.
17. Warren, G., Shklovski, I., Augenstein, I.: Show Me the Work: Fact-Checkers' Requirements for Explainable Automated Fact-Checking. Conference on Human Factors in Computing Systems - Proceedings . (2025). https://doi.org/10.1145/3706598.3713277/SUPPL_FILE/PN8110-TALK-VIDEO-CAPTION.VTT.
18. Gaspar-Figueiredo, D., Fernández-Diego, M., Abrahão, S., Insfran, E.: Integrating Human Feedback into a Reinforcement Learning-Based Framework for Adaptive User Interfaces. Proceedings of The 29th International Conference on Evaluation and Assessment in Software Engineering (EASE 2025). 1, (2025). https://doi.org/https://doi.org/10.48550/arXiv.2504.20782.
19. Ng, C., Fung, Y.: Educational Personalized Learning Path Planning with Large Language Models. (2024). https://doi.org/https://doi.org/10.48550/arXiv.2407.11773.
20. Petridis, S., Terry, M., Cai, C.J.: PromptInfuser: Bringing User Interface Mock-ups to Life with Large Language Models. Extended Abstracts of the 2023 CHI Conference on Human Factors in Computing Systems (CHI EA '23), April 23â•fi28, 2023, Hamburg, Germany. 1, (2023). https://doi.org/10.1145/3544549.3585628.
21. Petridis, S., Terry, M., Cai, C.J.: PromptInfuser: How Tightly Coupling AI and UI Design Impacts Designers' Workflows. Proceedings of the 2024 ACM Designing Interactive Systems Conference, DIS 2024. 1, 743–756 (2023). https://doi.org/10.1145/3643834.3661613.
22. Cornelis, L., Bernárdez, G., Jeong, H., Miolane, N.: When Machine Learning Gets Personal: Understanding Fairness of Personalized Models.
23. Schäfer, R., Sahabi, S., Preuschoff, P., Borchers, J.: Leveraging Digital Accessibility Using Generative AI. Proceedings of. 1,. https://doi.org/10.17605/osf.io/tgrw9.
24. Wu, J., Schoop, E., Leung, A., Barik, T., Bigham, J.P., Nichols, J.: UICoder: Finetuning Large Language Models to Generate User Interface Code through Automated Feedback. https://doi.org/https://doi.org/10.48550/arXiv.2406.07739.
25. Wei, J., Courbis, A.-L., Lambolais, T., Dray, G., Maalej, W.: On AI-Inspired UI-Design. (2024). https://doi.org/https://doi.org/10.48550/arXiv.2406.13631.
26. HULAT Homepage, https://access2meet.uc3m.es/resultados-2/, last accessed 2025/11/08.
27. Martínez, P., Moreno, L., Ochoa, H., Ramos, A., Pérez-Enríquez, M.: A Tool Suite for Cognitive Accessibility Leveraging Easy-to-Read Resources and Simplification Strategies. (2024).